\documentclass[twocolumn,showpacs,preprintnumbers,amsmath,amssymb]{revtex4}
\usepackage{graphicx} % Include figure files
\usepackage{dcolumn} % Align table columns on decimal point
\usepackage{bm} % bold math

\begin{document}

\preprint{APS/xxx-xxx}

\title{Non-equilibrium Kondo effect in asymmetrically coupled quantum dot}

\author{M. Krawiec}
\author{K. I. Wysoki\'{n}ski}
\affiliation{Institute of Physics, M. Curie-Sk\l odowska University, 
             ul. Radziszewskiego 10A, 20-031 Lublin, Poland}

\date{\today}

\begin{abstract}
The quantum dot asymmetrically coupled to the external leads has been analysed
theoretically by means of the equation of motion ($EOM$) technique and  the
non-crossing approximation ($NCA$). The system has been described by the 
single impurity Anderson model. To calculate the conductance across the device 
the non-equilibrium Green's function technique has been used. The obtained 
results show the importance of the asymmetry of the coupling for the 
appearance of the Kondo peak at nonzero voltages and qualitatively explain 
recent experiments. 
\end{abstract}


\maketitle

%%%%%%%%%%%%%%%%%%%%%%%%%%%%%%%%%%%%%%%%%%%%%%%%%%%%%%%%%%%%%%%%%%%%%%%%%%%%%%

\section{\label{sec1} Introduction}

Recent advances in nanotechnology have allowed the fabrication of structures 
containing quantum dots coupled to the external environment. 
The quantum dot consists 
of finite number of electrons confined to the small region of space. 
It behaves 
like an impurity in a metal \cite{PALee} and allows the study of the many body 
correlations between electrons. However, unlike an impurity which parameters 
are fixed, the coupling of the quantum dot to the external leads and its other 
parameters can be changed in a highly controlled way. Most importantly the 
non-equilibrium transport \cite{non-eq} can also be studied.

The discovery \cite{GGordon,SaraCr} of the Kondo effect \cite{Hewson}  in 
the quantum dots connected to external leads 
by tunnel junctions has resulted in an 
increased experimental \cite{GGordon2}-\cite{vKlitzing2} and theoretical 
\cite{multilevel}-\cite{time-dep} interest in this many body phenomenon. 
The Kondo effect in the quantum dot manifests 
itself at temperatures $T$ lower than 
the Kondo temperature $T_K$ as an increased conductance $G$ through the system. 
It is due to the formation of the so called 
Abrikosov-Suhl or Kondo resonance at 
the Fermi energy. This is a many body singlet state involving spin on the 
the quantum dot and the electrons in external leads.

The experiments \cite{GGordon,SaraCr,GGordon2,Wiel} have confirmed the validity 
of the theoretical picture. They also  discovered  phenomena 
 the explanation of which 
requires new  theoretical ideas. 
In particular they have shown the non-orthodox 
and unexpected behavior of the systems in the Kondo regime. These are 
{\it inter alia} the observation of the Kondo peak at nonzero source-drain 
voltage \cite{vKlitzing,Simmel}, absence of the odd-even parity effects 
expected for these systems \cite{vKlitzing2} and observation of the 
singlet-triplet transition in a magnetic field. Besides the non-linear 
current-voltage characteristics it has been possible to measure  charge 
distribution which led to the conclusion of spin-charge separation in a Kondo 
regime, observe the evolution of the transmission phase \cite{Ji} and the 
detection of two different energy scales \cite{Wiel2} related to two stages of 
the spin screening process in systems with spin $S\ge 1$, with one of the Kondo 
temperatures as high as  $4K$. 

The great progress in theoretical understanding of the Kondo physics in real 
quantum dots has been made during last decade. The theory has concentrated on 
such important aspects as the Kondo-driven transport in multilevel quantum dots 
\cite{multilevel}, the coupled quantum dots \cite{Ivanov}, double-dot 
structures in which existence of the Kondo effect without spin-degree of 
freedom and new singlet-triplet effects have been predicted \cite{doubledot}, 
the nature (weak {\it vs} strong coupling) of the Kondo effect at high voltage 
\cite{Kaminski}, the spin-charge separation in the strongly correlated quantum 
dot \cite{spin-ch}, the systems driven out of equilibrium by different means 
\cite{time-dep} {\it etc.}.

Here we shall focus our attention on the experimental observation 
\cite{vKlitzing,Simmel} of the Kondo effect at nonzero source drain voltages. 
To state the problem in the right perspective let us remind that in  systems 
containing quantum dot, the Kondo effect manifests itself at low temperatures 
as an enhanced conductance observed at zero source-drain voltage, 
$V_{\rm SD}=0$ \cite{PALee,non-eq,Hewson}. Occasionally, the Kondo 
peak in conductance 
appearing at nonzero voltages $V_{\rm SD}\neq 0$ \cite{vKlitzing} has been 
observed and this unusual behavior, called anomalous Kondo effect remains 
unexplained. Recently this phenomenon  has been studied systematically 
\cite{Simmel}. The authors have fabricated 
the dot coupled weakly to one and strongly to another lead and observed the 
evolution of the peak at $V_{{\rm SD}}\neq 0$. The source-drain voltage 
$V_{{\rm SD}}$,  at which the peak  appears, scales roughly linearly with a 
gate voltage, $V_{g}$. In the experimental setup \cite{Simmel} the 
additional electrode  determines the asymmetry in the coupling of the dot 
to left and right leads. 

It is the purpose of this paper to study the anomalous 
Kondo peak observed at non zero 
voltage. We shall present the results of the model calculations based on the 
non-equilibrium transport theory \cite{Keldysh} applied to the quantum dot 
described by the Anderson model \cite{Anderson} with asymmetric coupling to the 
leads. As we shall see the asymmetry in the couplings is the main factor which 
leads to this anomalous Kondo effect. The experimentally observed shifts of the 
Kondo peak to higher values of $V_{\rm SD}$ with increasing gate voltage can be 
satisfactorily explained by assuming that the values of the left and right 
barriers change together with the gate voltage, while the asymmetry in the 
couplings remains constant. This scenario is realized in experiment 
\cite{Simmel}.
 
The organization of the rest of the paper is as follows. In section \ref{sec2} 
we introduce the model, give the formula for the current through the quantum 
dot and discuss briefly the methods (equation of motion ($EOM$) with slave 
boson representation of electron operators and non-crossing approximation 
($NCA$)) used to calculate on-dot Green's function  relegating 
some technical details
 to the appendix. In section \ref{sec3} we 
present the results of our numerical calculations of the tunneling conductance 
across the asymmetrically coupled single level quantum dot in $U=\infty$ limit. 
Conclusions are given in section \ref{sec4}.

%%%%%%%%%%%%%%%%%%%%%%%%%%%%%%%%%%%%%%%%%%%%%%%%%%%%%%%%%%%%%%%%%%%%%%%%%%%%%%

\section{\label{sec2} The theory}

For the sake of simplicity we discuss here the dot with single energy level. 
The theoretical analysis of the transport through quantum dot usually starts 
with the following, Landauer type, formula \cite{non-eq} for the current
\begin{eqnarray}
J ={\frac{e}{\hbar}}\sum_\sigma \int d\omega [f_L(\omega) -
f_R(\omega)] 
\nonumber \\
\times {\frac{\Gamma^L_\sigma(\omega)\Gamma^R_\sigma(\omega) }
{\Gamma^L_\sigma(\omega) + \Gamma^R_\sigma(\omega)}} 
{\left(\frac{- 1}{\pi}\right)} {\rm Im} G_{\sigma}^r(\omega+ i0^+)
\label{curr}
\end{eqnarray}
Here $f_\lambda(\omega)$ denotes the Fermi distribution function for lead 
$\lambda $ with chemical potential $\mu _{\lambda }$, $G_{\sigma }^r(\omega )$ 
is the (retarded) impurity Green's function and 
$\Gamma ^{\lambda }_{\sigma}(\omega )=2\pi\sum_{k}|V_{\lambda k}|^{2} 
 \delta(\omega -\varepsilon _{\lambda k})$ is the effective coupling of 
localized electron to conduction band.

The current $J$ flowing across the system depends on the source-drain voltage 
$V_{SD} = (\mu_L - \mu_R)/e$, where $e$ is the electron charge. The 
differential conductance of the system defined as 
$G(V_{SD}) = \frac{dJ(V_{SD})}{dV_{SD}}$ is directly measured experimentally 
\cite{GGordon}.

To calculate  the on-dot Green's function $G_{\sigma }(\omega +i0^{+})$ we 
shall describe the dot coupled to the external leads by the single impurity 
Anderson Hamiltonian \cite{Anderson} 
\begin{eqnarray}
H = \sum_{\lambda k\sigma }\varepsilon _{\lambda k} 
c_{\lambda k\sigma}^{+} c_{\lambda k\sigma }+
E_{d}\sum_{\sigma }d_{\sigma }^{+}d_{\sigma}
\nonumber \\
+ Un_{\uparrow }n_{\downarrow }+ 
\sum_{\lambda k\sigma }(V_{\lambda k} 
c_{\lambda k\sigma}^{+}d_{\sigma}+H.c.)
\end{eqnarray}
Here $\lambda =R,L$ denote the right ($R$) or the left ($L$) 
lead in the system. 
Other symbols have the following meaning: $c_{\lambda k\sigma}^{+}$ 
($c_{\lambda k\sigma }$) denotes creation (annihilation) operator for a 
conduction electron with wave vector $\vec{k}$, spin $\sigma $ in the lead 
$\lambda $, $V_{\lambda k}$ is the hybridization matrix element between 
conduction electron of energy $\varepsilon_{\lambda k}$ in the
 lead $\lambda$ and 
localized electron on the dot. $E_{d}$ is the single particle energy at the 
dot.  $n_{\uparrow }=d_{\uparrow}^{+}d_{\uparrow }$ is the number operator 
for electrons with spin up localized on the dot and $U$ is the (repulsive) 
interaction energy between two electrons. Our calculations are restricted
to very low temperatures, much smaller than the orbital level
spacing in quantum dot so it is legitimate to consider single 
energy level $E_d$.

There are various methods \cite{non-eq} of calculating 
the on-dot Green's function entering 
the current (\ref{curr}). Here we shall apply two of them: equation of 
motion method ($EOM$) and non-crossing approximation ($NCA$). 
In both cases we assume that the Coulomb repulsion $U$ between electrons
on  the dot is the largest energy scale. Therefore we take the limit 
 $U=\infty $. The
original correlated electron operators are expressed as
products  of
auxiliary fermion and boson ones \cite{non-eq}. 

 When  using equation of 
motion method ($EOM$) we apply a mean field like
approximation for the slave bosons and  calculate
all matrix elements of the Keldysh Green's functions, including the
distribution one \cite{EOM}. 
 In the process we consistently
decouple all elements of 
the higher order Keldysh Green functions \cite{MAK-KIW}.
The relevant formulae and some technical details can be found in the appendix.
 As we shall see
the method gives correct position of the Kondo peak. 
 However, like the standard $EOM$ it leads to 
 incorrect width of the peak and the occupations. Therefore we have 
used  the non-crossing approximation ($NCA$), which is generally accepted 
technique of solving the problem at hand \cite{Hewson}.
In the $NCA$ one maps the infinite $U$ Anderson model onto
the slave boson one and calculates both boson 
and fermion propagators. They are expressed by the coupled 
integral equations \cite{non-eq}.

 Where appropriate we 
shall present the results obtained by both techniques.
 
%%%%%%%%%%%%%%%%%%%%%%%%%%%%%%%%%%%%%%%%%%%%%%%%%%%%%%%%%%%%%%%%%%%%%%%%%%%%%%

\section{\label{sec3} Numerical results}

Let us  first discuss the relation between the experimental parameters and 
those entering the model and the theory. The effective coupling $\Gamma_L$ and
$\Gamma_R$ have been estimated in ref.\cite{Simmel} to be $170 \; \mu eV$ and
$80 \; \mu eV$ respectively. Their values and the ratio 
$\Gamma_L/\Gamma_R \approx 2$ have been argued to remain constant during the 
measurements. The source-drain voltage $V_{SD}$ is  the difference of the 
chemical potentials of the external leads. The (back)gate voltage $V_g$ 
controls the position of the on-dot energy level $E_d$. As already mentioned we 
stick here to the $U=\infty $ limit. In this limit there can be at most a 
single electron with energy $E_d$ on the dot at a time. 

\begin{figure}[h]
 \resizebox{8.4cm}{!}{
  \includegraphics{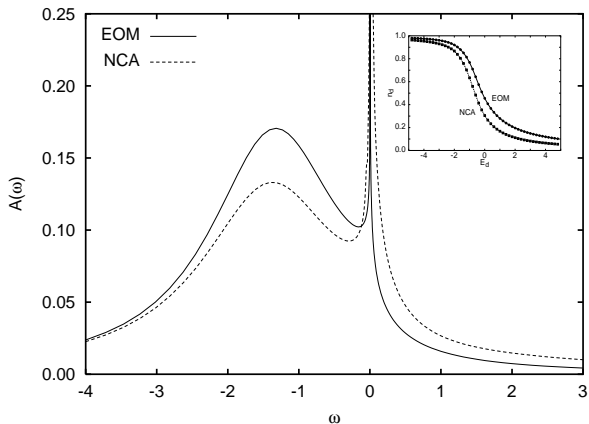}}
 \caption{\label{Fig1} The equilibrium density of states on the quantum dot 
          obtained within $EOM$ (solid line) and $NCA$ (dashed line). Note the 
	  relative shift of the spectral weight with respect to the chemical 
	  potential $\mu=0$ which results in different occupations shown in the 
	  inset. The parameters are: $E_d=-2$, $\Gamma_L = \Gamma_R = 1$ and 
	  $T=10^{-3}$.}
\end{figure}

We start the presentation of the results with the comparison of the
(equilibrium) density of states of a quantum dot coupled to two leads obtained 
by means of $NCA$ and $EOM$ approaches. It is shown in the Fig. (\ref{Fig1}).
The main features of the $DOS$ remain the same in both approaches. 
However the
height and the width of the Kondo peak is much larger in the $NCA$. 
Moreover the spectral
weight  is shifted towards higher energies. In turn this
leads to different occupations shown in the inset of the Fig.(\ref{Fig1}).

\begin{figure}[h]
 \resizebox{8.4cm}{!}{
  \includegraphics{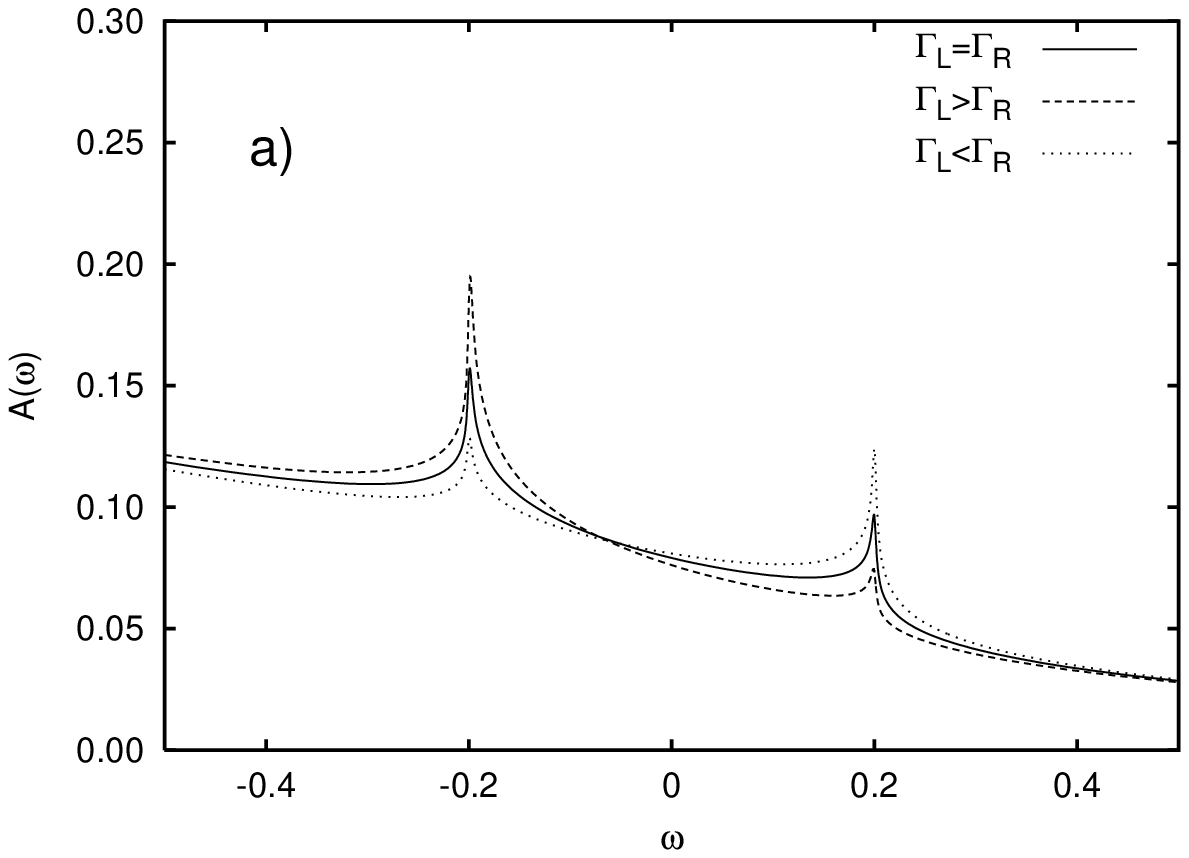}}
 \resizebox{8.4cm}{!}{
  \includegraphics{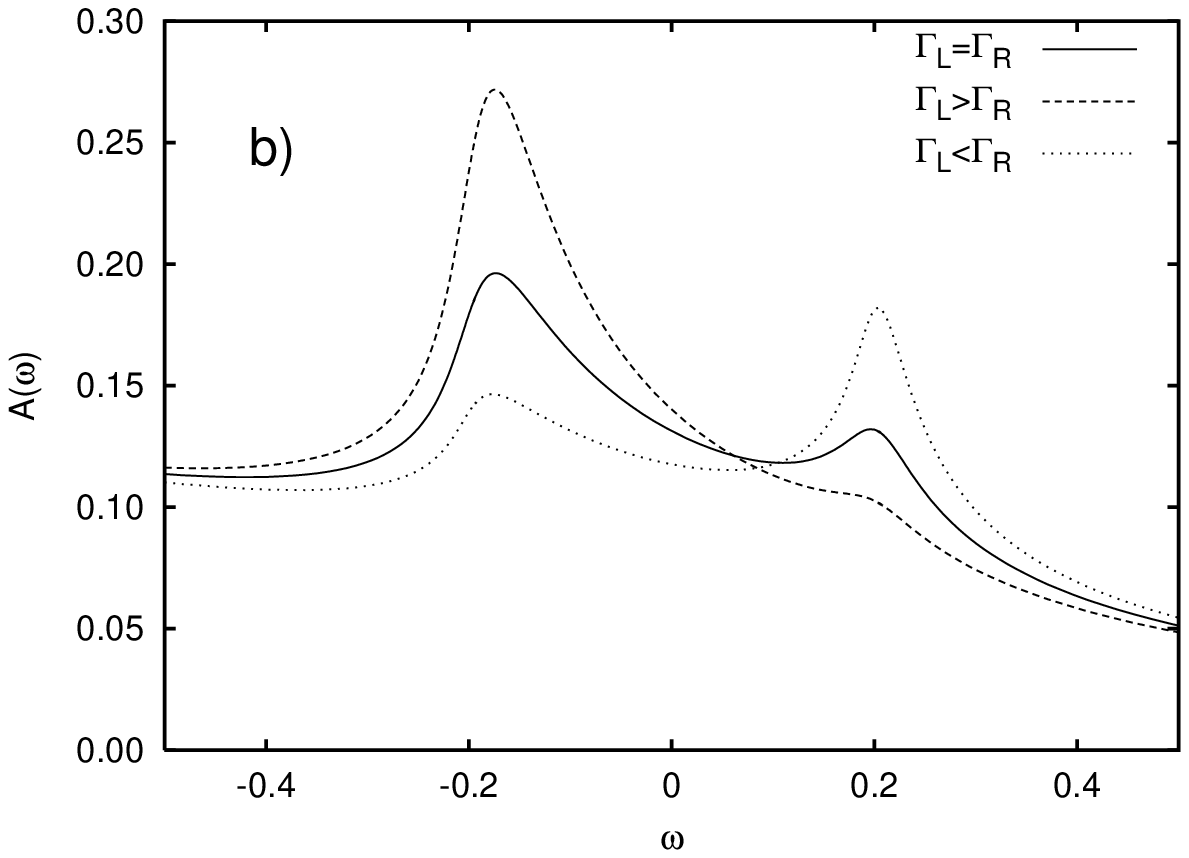}}
 \caption{\label{Fig2} The nonequilibrium density of states obtained within
          $a)$ - $EOM$ and $b)$ - $NCA$ for the symmetric 
	  $\Gamma_{L} = \Gamma_{R}$ (solid lines) and asymmetrically coupled 
	  quantum dot with $\Gamma_{L} = 2 \Gamma_{R}$ (dashed) and 
	  $\Gamma_L = \frac{1}{2} \Gamma_R$ (dotted lines). 
	  $\mu_R = - \mu_L = 0.2$ and the other parameters are the same as in 
	  Fig. (\ref{Fig1}).}
\end{figure}

Now let's turn to the nonequilibrium ($\mu_L \neq \mu_R$) density of states. In
this case the high energy features to large extend 
remain the same as in equilibrium (see
Fig.(\ref{Fig1})), so the only low energy $DOS$ is shown in the 
Fig.(\ref{Fig2}).
The upper panel presents results obtained {\it via} equation of motion 
technique for Keldysh (matrix) Green's function.In the lower panel the 
results obtained with the non-crossing approximation are shown. The coupling is 
asymmetric  with $\Gamma_L /\Gamma_R = 2$ (dashed lines) and $\frac{1}{2}$ 
(dotted lines). The case of the symmetric coupling $\Gamma_L = \Gamma_R$ is 
also shown (solid lines) for comparison. 
Few features have to be noted. First we see that the 
Kondo peak is always located at energies coinciding with those of the  Fermi 
levels of the leads. Thus in non-equilibrium  we get (in the density of states) 
two Kondo resonances pinned to Fermi energies of the left and right 
electrodes. Note also that the heights  
of the respective Kondo resonances 
strongly depend on the value of the hybridization. The overall shape of the 
density of states is similar. The positions of the Kondo peaks are roughly the 
same but they differ in width and heigths. 
The peaks obtained in $EOM$, are much 
narrower and smaller. As a result the curves in figure (\ref{Fig2}a) 
differ from that in figure (\ref{Fig2}b). 
   
   These details in the energy dependence of the density of states 
   may shortly be a matter of 
direct measurements. In fact it has been recently predicted theoretically 
\cite{Lebanon} that the on-dot density of states can be measured in a device 
containing quantum dot coupled to three leads. The very weakly coupled third 
lead will act as a tunneling tip
in conventional tunneling microscope and will  
probe the non-equilibrium density of states. The conductance spectrum 
measured by this additional electrode has been shown \cite{Lebanon} to
follow the non equilibrium density of states, like one shown in Fig.(2).

\begin{figure}[h]
 \resizebox{8.4cm}{!}{
  \includegraphics{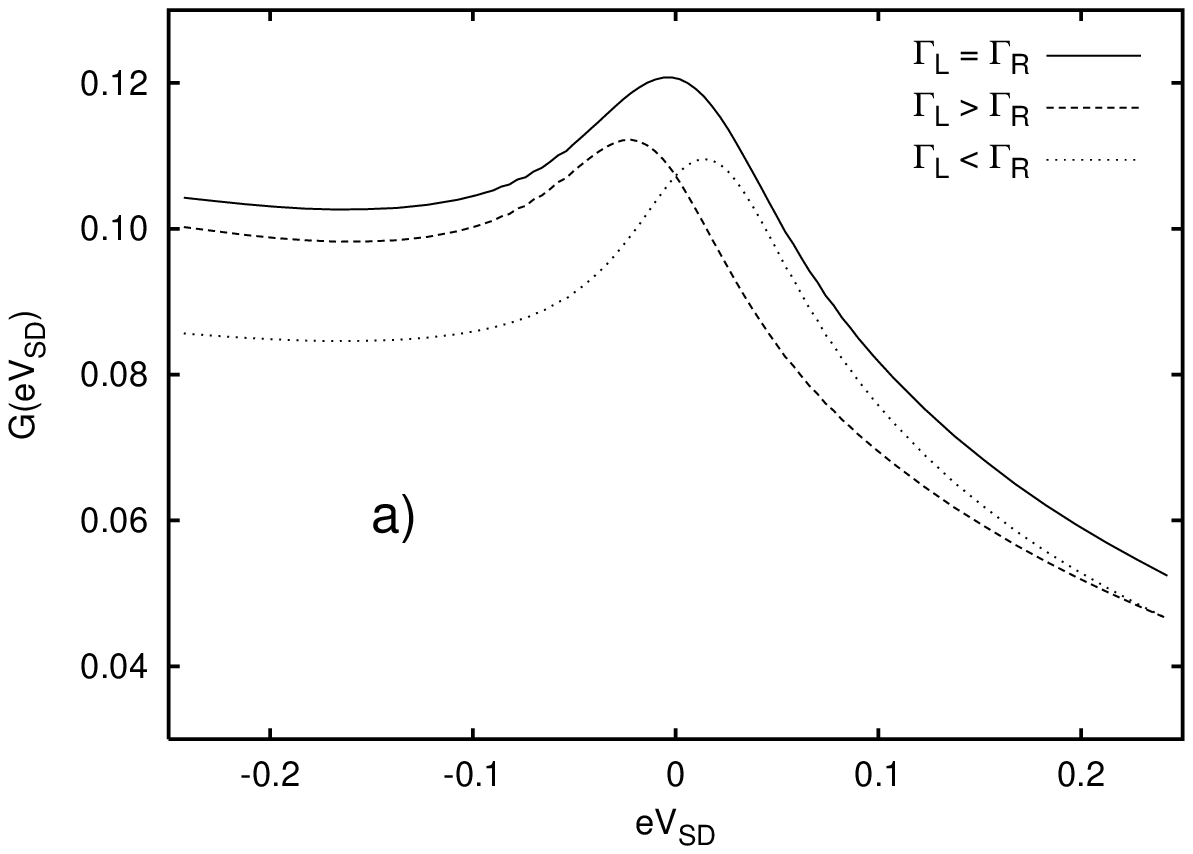}}
 \resizebox{8.4cm}{!}{
  \includegraphics{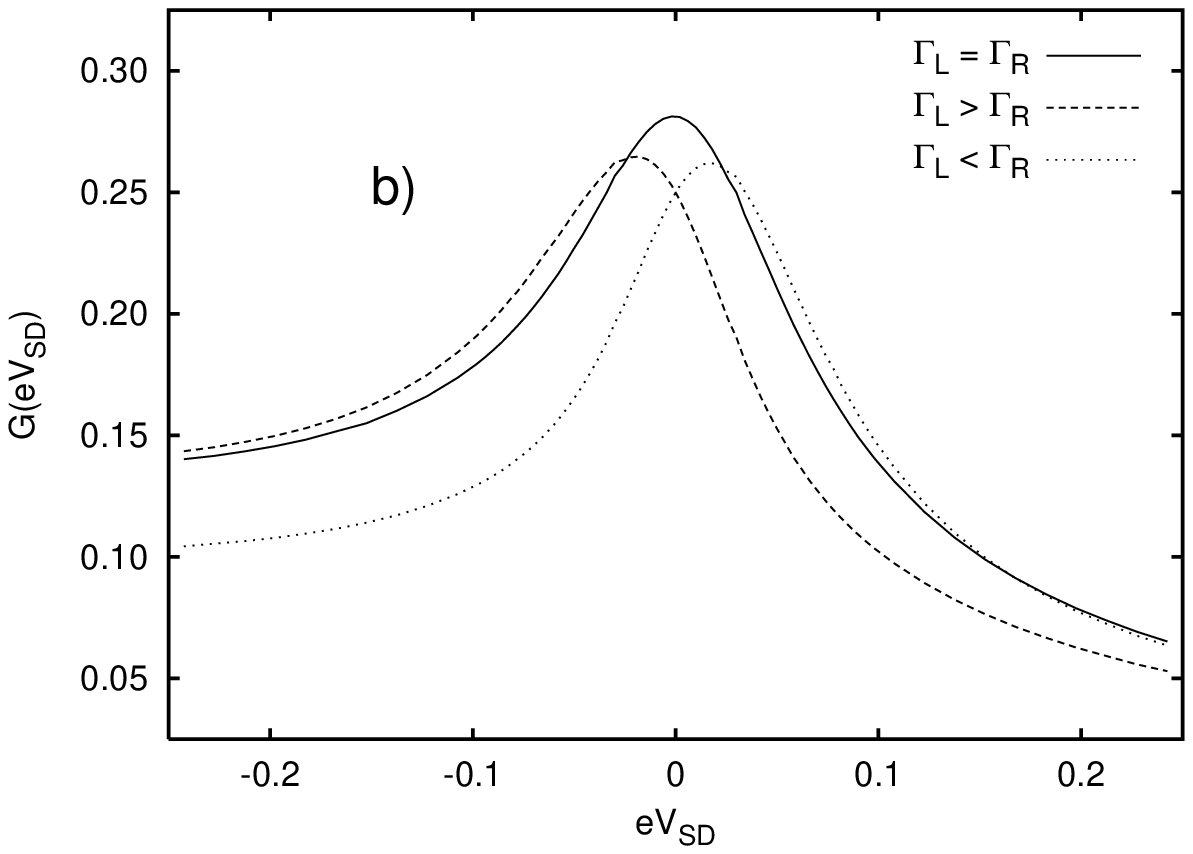}}
 \caption{\label{Fig3} The differential conductance 
          ($G(eV_{SD}) = d J/d (eV_{SD}$)) obtained within
          $a)$ - $EOM$ and $b)$ - $NCA$ for the symmetric 
	  $\Gamma_{L} = \Gamma_{R}$ (solid lines) and asymmetrically coupled 
	  quantum dot with $\Gamma_{L} = 2 \Gamma_{R}$ (dashed) and 
	  $\Gamma_L = \frac{1}{2} \Gamma_R$ (dotted lines).}
\end{figure}

Returning to our main subject we show in Fig.(\ref{Fig3}) the
differential conductance spectrum corresponding to the same 'experimental 
setup' as discussed previously in connection with  figure (\ref{Fig2}). 
For comparison we  have also plotted in this figure the conductance through the 
symmetrically coupled quantum dot. In the symmetric situation 
($\Gamma_L = \Gamma_R$) the Kondo resonance is located exactly at zero bias 
($V_{DS}=0$), but for $\Gamma_L > \Gamma_R$ ($\Gamma_L < \Gamma_R$) it is 
shifted to the negative (positive) voltages $V_{DS}$. This finding is in nice 
qualitative agreement with the experimental studies on the transport through 
the quantum dot in the presence of the asymmetric barriers \cite{Simmel}. 
 While the observed shifts calculated within $EOM$ and $NCA$ are of comparable
 magnitude the clear differences in their shape are visible. 
 The $NCA$ peaks are
 much higher and more symmetric in vicinity of  their maxima.
 For asymmetric coupling the Kondo resonance 
in the conductance is pinned to the position of the
Fermi level of that lead which is more strongly coupled 
(larger $\Gamma$) to the dot. It is thus mainly the relative coupling
which rules the value of the shift.

\begin{figure}[h]
 \resizebox{8.4cm}{!}{
  \includegraphics{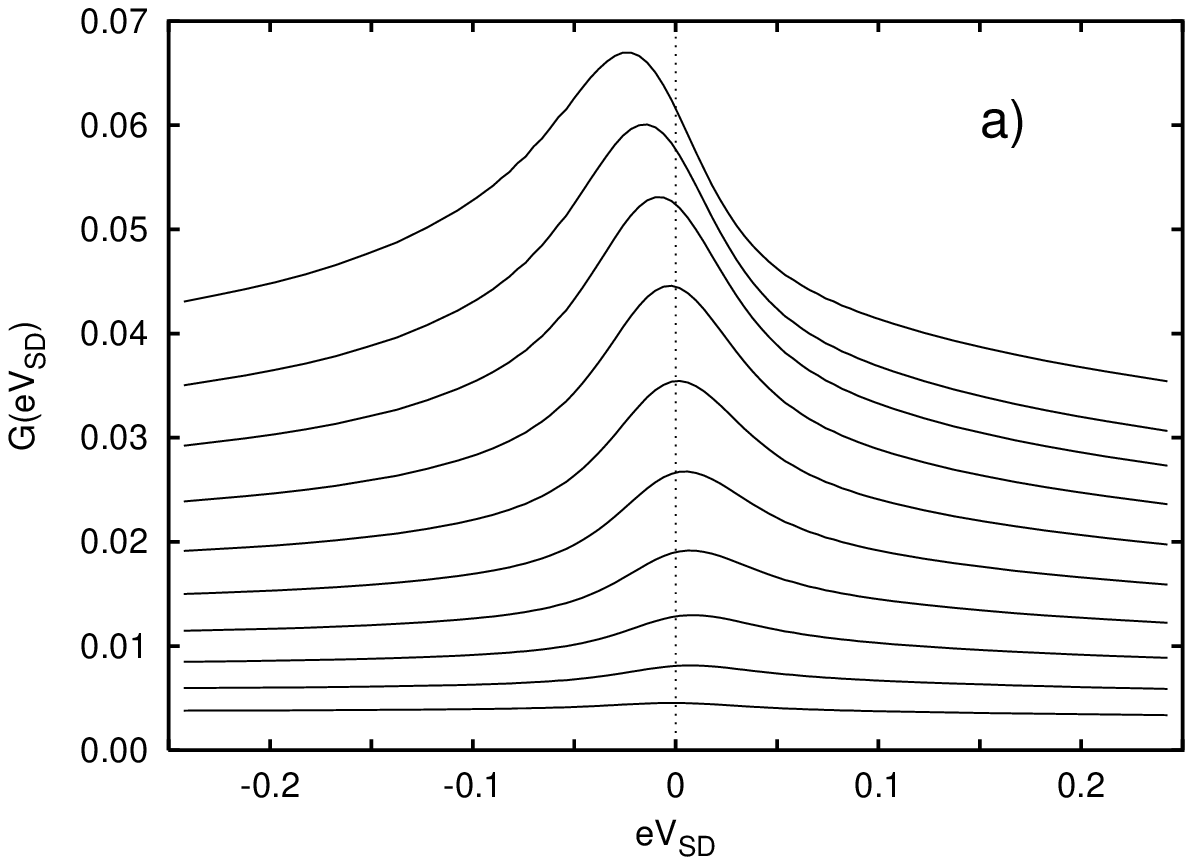}}
 \resizebox{8.4cm}{!}{
  \includegraphics{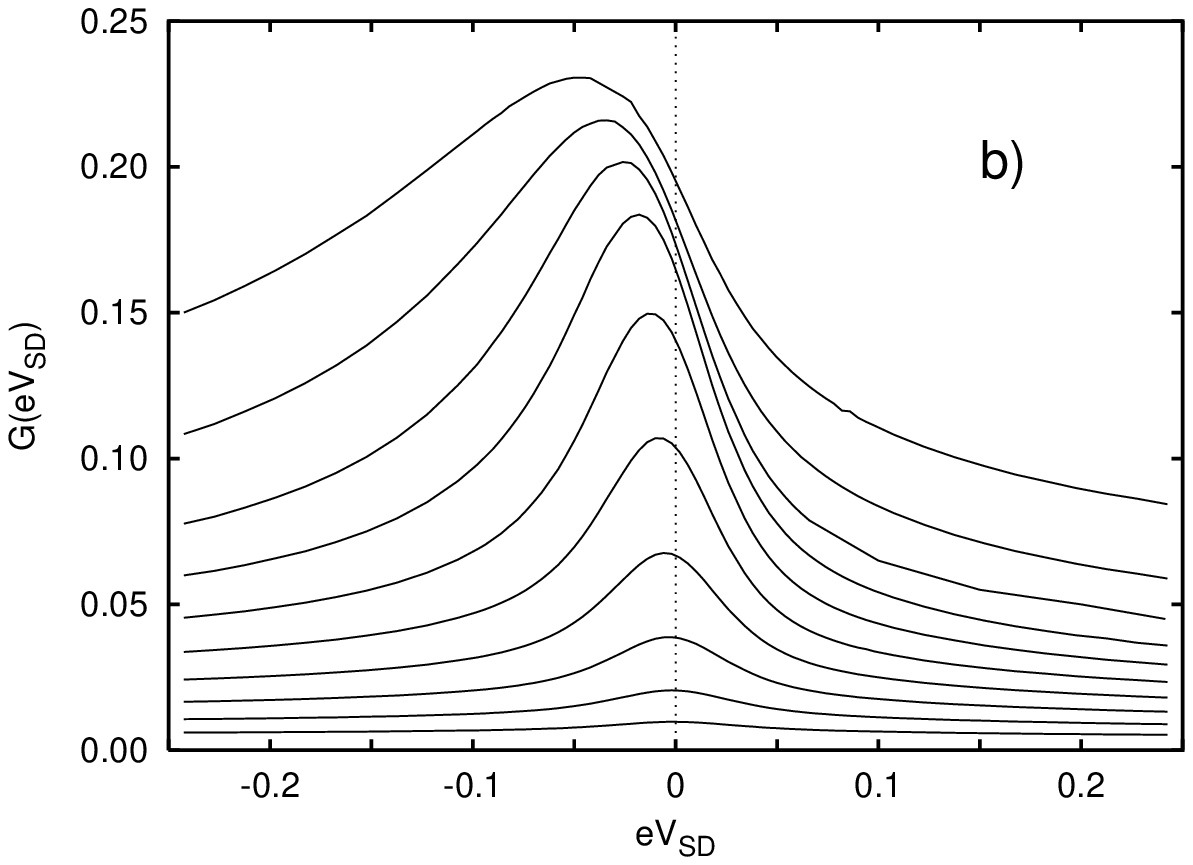}}
 \caption{\label{Fig4} The differential conductance obtained within
          $a)$ - $EOM$ and $b)$ - $NCA$ for the different asymmetric couplings.
	  The lower curve is for $\Gamma_{L} = \Gamma_{R}$ while the upper one 
	  - $\Gamma_L = 5.5 \Gamma_R$. Other parameters are $E_d = -10$ and 
	  $T= 10^{-2}$ in units of $\Gamma_R$.}
\end{figure}

In  Fig.(\ref{Fig4}) we show the systematic change of the $G(V_{DS})$ 
with increasing asymmetry $\Gamma_L / \Gamma_R$ of the coupling. 
The upper curves in both panels corresponds  to $\Gamma_L/\Gamma_R=5.5$  while 
lower one is for symmetric coupling $\Gamma_L/\Gamma_R=1$ in steps of 0.5.
 The increase of the 
asymmetry $\Gamma_L/\Gamma_R$ from $1$ to $5.5$ continuously moves the Kondo 
peak away from $V_{SD} = 0$ position. We have checked that increasing asymmetry 
to still higher values does not lead to bigger shifts. This is easy to 
understand as for large asymmetry one of barriers is 
not transparent enough to 
produce the clear Kondo resonance in the density of states.

  The position of the on-dot electron energy level $E_d$ 
  influences  anomalous Kondo peak for the asymmetric dot with 
$\Gamma_L/\Gamma_R=2$  to lesser extend. It is only 
important that it takes a value appropriate for observing 
a Kondo resonance. For all appropriate $E_d$ the shifts
are of comparable magnitudes. 

The data displayed in  the figure (\ref{Fig4})   qualitatively 
agree with those plotted in figure (5) of \cite{vKlitzing} and figure (3) of 
\cite{Simmel}. However, theoretical shifts of the Kondo peak position  are 
smaller than the experimental. 

There may be additional factors which affect the position of the peaks. We have 
checked that the energy dependence of $\Gamma_{L,R}$ introduces only small 
quantitative differences in the density of states and differential conductance, 
and does not lead to better agreement between theory and experiment. Similarly 
the calculations within $EOM$ approach for finite values of $U$ show that 
finite $U$ leads to minor 
corrections as also does the presence of the additional energy levels in the 
vicinity of Fermi energy. In all the cases studied one gets 
usual behavior with Kondo 
peak located at $V_{SD}=0$ for symmetric coupling to both leads and the 
anomalous Kondo effect for asymmetric coupling. This proves the importance of
the asymmetry in the observattion of it.

In experimental setup \cite{Simmel} the changes of the gate 
voltage $V_g$, which in first place affect the 
position of the electron energy level  also modify the height of the 
barriers and their transparency $V_{\bf k}$. This effect is of special 
importance in the quantum dots defined in the two dimensional electron gas 
where the voltage at a single electrode couples capacitively to other 
electrodes \cite{Buttiker}. If we assume that (as in experiments) 
$\Gamma_L/\Gamma_R$ remains constant ($=2$) and that the decrease of the 
energy $E_d$ is accompanied by the simultaneous increase of the couplings 
$\Gamma_R$ and $\Gamma_L$ then the calculated shifts get larger.

The occurence of the Kondo resonace  is possible at
low enough temperature. It is also well known \cite{non-eq} 
that changes of temperature  move slightly the  Kondo peak. 
We have checked 
this  and found that if temperature is raised 
the position of the peak moves slightly away 
from the $V_{SD}=0$. At finite temperature the occupation of
the dot changes and the Abrikosov-Suhl resonances smear out
 and this leads to
small changes in the position of the Kondo peak.

We thus have combined all above contributions, i.e. 
assymetry in the couplings $\Gamma_L \neq \Gamma_R$, their $E_d$-dependence, 
and assumed high enough, but still below $T_K$, temperature to get larger 
shifts of the Kondo peak. 
\begin{figure}[h]
 \resizebox{8.4cm}{!}{
  \includegraphics{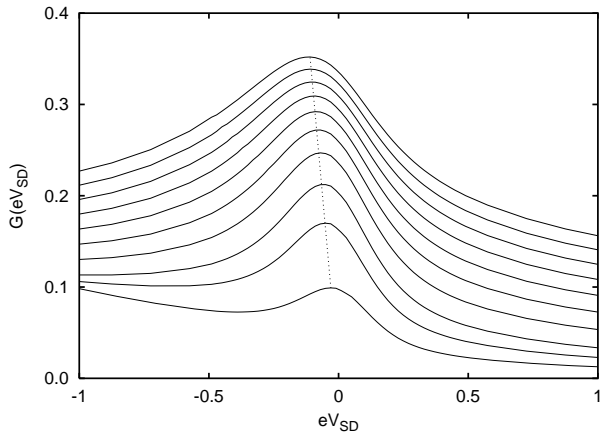}}
 \caption{\label{Fig5} The $NCA$ calculated differential 
          conductance as a function of 
          source-drain voltage $V_{SD}$ for various values of the $E_d$ and 
	  $\Gamma_L$, $\Gamma_R$ at fixed $\Gamma_L/\Gamma_R = 2$. The lowest 
	  curve correspond to $E_d = -3$, $\Gamma_L = 1$, while upper one is 
	  for $E_d = -12$, $\Gamma_L = 6.4$ in units of $\Gamma_0 $ equal to
	  initial coupling of the left lead. The
	  temperature $T = 5 \; 10^{-2}$ is below estimated Kondo
	   temperatures. }
\end{figure}
 We have shown the results in Fig.(\ref{Fig5}). The various
  curves have been calculated for  $T=5 10^3$ which is 
   below Kondo temperature. 
In the figure the change of the 
position of the on dot energy level is accompanied 
by the simultaneous change 
of the barrier transparency.  The bottom 
curve in  figure (\ref{Fig5})corresponds to $E_d=-3\Gamma_0$, 
$\Gamma_0=\Gamma_L$, while the upper 
one corresponds to $E_d = -12\Gamma_0$, $\Gamma_L = 6.4\Gamma_0$. Here 
$\Gamma_0$ is equal to experimentally estimated value of
of the smaller of couplings \cite{Simmel}. The ratio 
$\Gamma_R/\Gamma_L$  is kept constant and equal $2$ as estimated in 
\cite{Simmel}. The data are in nice qualitative agreement with experiments. 
The 
theoretical shifts, however, are smaller than experimental by a factor of 
$5$-$10$. To check whether this is due to different asymmetry ratio we plot in 
figure (\ref{Fig6}) the results obtained for $\Gamma_R/\Gamma_L=4$.
 The  shifts  have increased.
\begin{figure}[h]
 \resizebox{8.4cm}{!}{
  \includegraphics{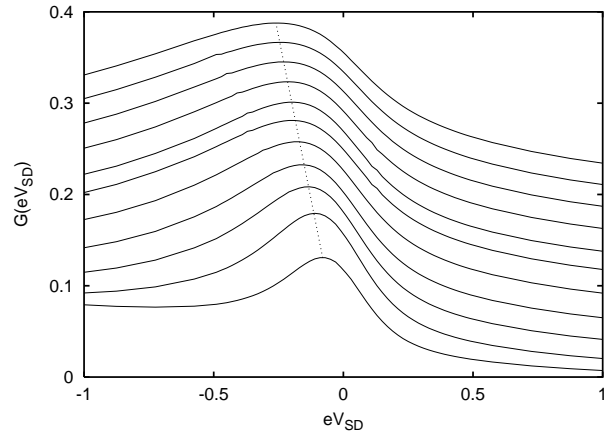}}
 \caption{\label{Fig6} The same as in Fig. (\ref{Fig5}) but for asymmetry 
          $\Gamma_L/\Gamma_R = 4$. The lowest curve correspond to $E_d = - 4$, 
	  $\Gamma_L = 1$, while upper one is for $E_d = - 17.5$, 
	  $\Gamma_L = 12.8$ in units of $\Gamma_0$ equal to initial coupling 
	  of the left lead.}
\end{figure}
%
  
%%%%%%%%%%%%%%%%%%%%%%%%%%%%%%%%%%%%% 

\section{\label{sec4} Conclusions}

We have found that the emergence of the Kondo peak at  non zero voltages 
$V_{DS}\neq 0$ is caused by asymmetric coupling of the dot to the external 
electrodes. These results are in qualitative agreement with experimental data 
on the transport through the quantum dot asymmetrically coupled to the leads 
\cite{vKlitzing, Simmel}. The theoretical Kondo peaks in differential 
conductance, however, are narrower than experimental ones. Their maxima move 
to nonzero $V_{SD}$  with increasing the asymmetry or the position of the 
on-dot energy level. The simultaneous change of $E_d$ and $\Gamma_L$, 
$\Gamma_R$ can semi-quantitatively explain experimental data. More experimental 
results are needed to draw firm conclusions such as the applicability of simple 
Anderson model to asymmetrically coupled quantum dots. It follows from the 
presented studies that the asymmetry of the couplings is a necessary ingredient 
for the explanation  of the anomalous Kondo effect. Within the Anderson model 
one always gets normal Kondo effect for symmetric couplings and small shifts of 
the Kondo peak to non-zero voltages for asymmetric couplings. Our inability to 
explain quantitatively the experimental data may indicate the necessity of much 
better theoretical treatment of the model or even better model for the 
description of these complicated systems. There is a possibility that the 
experimentally observed features, even though similar to, do not represent 
genuine Kondo effect. In fact some researches \cite{DGG-privat} have seen very 
small shifts, consistent with present calculations, even for quite 
asymmetrically coupled quantum dots.

%%%%%%%%%%%%%%%%%%%%%%%%%%%%%%%%%%%%%%%%%%%%%%%%%%%%%%%%%%%%%%%%%%%%%%%%%%%%%%
 
\begin{acknowledgments}
This work has been partially supported by the State Committee for Scientific 
Research under grant 2P03B 106 18. We thank unknown Referee for her/his 
comments on the first version of the paper.
\end{acknowledgments}

%%%%%%%%%%%%%%%%%%%%%%%%%%%%%%%%%%%%%%%%%%%%%%%%%%%%%%%%%%%%%%%%%%%%%%%%%%%%%%

\appendix

\section{}

To find  the current accross the system, Eq.(\ref{curr}), it is enough 
to calculate the on-dot retarded Green's function.

The $NCA$ method to get $G^r_{\sigma}(\omega)$ has been extensively
discussed previously \cite{non-eq} and there is no need to 
repeat its derivation again. For the sake of completeness let us only 
note that we
have adapted the formulae derived in the second paper of the reference
\cite{non-eq}.

The $EOM$ method to calculate the $GF$ is straightforward and in 
$U\rightarrow \infty$ limit leads to 
\begin{eqnarray} 
G^r_{\sigma}(\omega)=\frac{1 - \langle n_{-\sigma} \rangle} 
{\omega - E_d -\sum_{\lambda} \Sigma^r_{\sigma\lambda} (\omega)} 
\label{GFinfty} 
\end{eqnarray} 
with the self-energy 
\begin{eqnarray} 
\Sigma^r_{\sigma\lambda} (\omega) = \sum_{\lambda {\bf k}}  
|V_{\lambda {\bf k}}|^2  
\frac{1+f (\omega-\mu_{\lambda})}{\omega - \epsilon_{\lambda {\bf k}}}. 
\label{Sigmainfty} 
\end{eqnarray}  

In  the equation (\ref{GFinfty}), $\langle n_{-\sigma} \rangle$ denotes the 
average on-dot occupation number of the spin $-\sigma$ electrons. In the 
equilibrium one calculates $\langle n_{-\sigma} \rangle$ self-consistently 
from the retarded Green's function $G^r_{-\sigma} (\omega)$. Here we are 
dealing with nonequilibrium situation and $\langle n_{-\sigma} \rangle$ cannot 
be calculated directly from $G^r_{-\sigma} (\omega)$. 
Instead the nonequilibrium  
\cite{Keldysh} Green's function technique has to be used. 
The occupation of the dot at time $t$ is expressed {\it via} Keldysh "lesser" 
Green's function  
$\langle n_{\sigma} (t) \rangle = \langle c^+_{\sigma} (t) \;  
c_{\sigma} (t) \rangle = -i G^<_{\sigma} (t,t)$. In the steady state  
one gets 
\begin{eqnarray}  
\langle n_{\sigma} \rangle = 
- i \int_{-\infty}^{\infty} \frac{d\omega}{2\pi} 
G^<_{\sigma} (\omega). 
\label{nSun}  
\end{eqnarray}  
This shows that the consistent calculations of the retarded $GF$ requires the 
knowledge of "lesser" one. The equation of motion for the "lesser"
 $GF$ has been 
formulated by Niu {\it et al} \cite{EOM}. For the Hamiltonian of the form 
$H=H_0+H_I$ 
they derived the following general equation   for the "lesser" GF
\begin{widetext}
\begin{equation}
\langle\langle A|B\rangle\rangle^<_{\omega} =
g^<(\omega)\langle[A,B]_{\pm}\rangle+g^r(\omega)
\langle\langle[A,H_I]|B\rangle\rangle^<_{\omega}+
g^<(\omega)\langle\langle[A,H_I]|B\rangle\rangle^a_{\omega},
\end{equation}
\end{widetext}
here $g^{<(r)}(\omega)$ is the "lesser" (retarded) $GF$ of the noninteracting 
part $H_0$ of Hamiltonian.

To treat strong correlations we use the version \cite{MAK-KIW} of  the 
slave boson 
technique and rewrite the Hamiltonian in the form
\begin{widetext}  
\begin{equation}
H^{SB} = \sum_{\lambda k\sigma}(\varepsilon_{\lambda
k}-\mu_{\lambda}) c^+_{\lambda k\sigma} c_{\lambda k\sigma} +
\varepsilon_d \sum_\sigma f^+_\sigma f_\sigma + \sum_{\lambda
k\sigma} V_{\lambda k} (c^+_{\lambda k\sigma} b^+f_{\sigma} +
f^+_\sigma b c_{\lambda k\sigma}),
\end{equation}
\end{widetext}
where new fermionic ($f^+_{\sigma},f_{\sigma})$ and bosonic ($b^+,b
)$ operators have been introduced.
 Calculating   the on-dot Green's function $G^<_{\sigma}(\omega) =
\langle\langle b^+f_\sigma |f^+_\sigma b\rangle\rangle^<_{\omega}$ 
we have taken the third term of $H^{SB}$  as an
interaction part $H_I$ and the first two terms of it as $H_0$.
  
The average occupation number is found to be
%
%\begin{widetext}
\begin{eqnarray} 
\langle n \rangle = -\frac{1}{2\pi} \sum_{\sigma} 
\int^{\infty}_{-\infty} d\omega \;  
 \sum_{\lambda}  
{\rm Im} \Sigma^r_{\sigma\lambda} (\omega)  f_{\lambda} 
(\omega) 
 |G_{\sigma}^r(\omega)|^2 . 
\label{ncorr} 
\end{eqnarray} 
%\end{widetext}
%
Note that in turn it depends  on the retarded Green's function. 
This closes the
system of equations.
%%%%%%%%%%%%%%%%%%%%%%%%%%%%%%%%%%%%%%%%%%%%%%%%%%%%%%%%%%%%%%%%%%%%%%%%%%   

\end{document}